\documentstyle[12pt,epsbox]{article}

\newcommand{\dlangle}{\left< \! \left<}
\newcommand{\drangle}{\right> \! \right>}
\newcommand{\erfc}{\mbox{{\rm erfc}}}

\begin{document}
\title{Retrieval Dynamics of Neural Networks for Sparsely Coded Sequential Patterns}
\author{\sc Katsunori Kitano and Toshio Aoyagi \\
\small \it Department of Applied Mathematics and Physics,\\
\small \it Graduate School of Informatics, Kyoto University, Kyoto 606}

\maketitle

\begin{abstract}
It is well known that 
a sparsely coded network in which the activity level is extremely low
has intriguing equilibrium properties.
In the present work,
we study the dynamical properties of a neural network
designed to store sparsely coded sequential patterns rather than static ones.
Applying the theory of statistical neurodynamics,
we derive the dynamical equations governing the retrieval process
which are described by some macroscopic order parameters such as the overlap.
It is found that our theory provides good predictions for the storage capacity and
the basin of attraction obtained through numerical simulations.
The results indicate that
the nature of the basin of attraction depends
on the methods of activity control employed.
Furthermore, it is found that
robustness against random synaptic dilution slightly deteriorates
with the degree of sparseness.
\end{abstract}

For the purpose of constructing more realistic mathematical neural network models (e.g, the Hopfield model~\cite{Hopfield}),
so-called ``random'' patterns, which have been used for simple theoretical treatments, have been reconsidered.
In a network capable of processing these random patterns,
it is frequently supposed that
statistically half of the neurons are allowed to be active.
However, such a situation is not realistic for two reasons.
First, according to the results of physiological studies,
the activity level of real neural systems is thought to be low.
Second, in a meaningful pattern, information is generally encoded 
by a small fraction of bits in a background 
which occupies most of the total area.
With these points in mind,
neural networks loading sparsely coded patterns have been studied by many authors~\cite{Willshaw,AGS,Tsodyks,Vicente,Horner,Buhmann}.
These authors have reported that
the maximal number of patterns stored in the network increases
as the fraction of active neurons $a$ decreases.
Furthermore, the storage capacity in such a situation diverges as $-1/a \ln a$
which is the optimal asymptotic form~\cite{Gardner}.
However, considering the fact that the information content in a single pattern
is reduced with the degree of sparseness,
we cannot immediately conclude that 
sparse coding enhances the associative ability.
Rather, what we should note is that 
the optimal bound is obtained even for models 
with a relatively simple Hebbian learning rule.

While, owing to these studies,
progress in the understanding of the equilibrium properties of sparsely coded networks has been made,
many unsolved problems remain in regard to dynamical aspects.
In order to grasp a network's characteristics properly,
it is necessary to consider the dynamical properties such as the basin of attraction.
In recent years, several theories treating retrieval process have been proposed~\cite{Rieger,HBFKS,GDM,Coolen,AmariMaginu,Okada}.
Among these, we note that 
the method of statistical neurodynamics is practically useful,
because it enables us to describe long-term behavior~\cite{AmariMaginu,Okada}.
However, for sparse coding,
there is quantitative discrepancy between the results obtained from this theory and numerical simulation in the case of autoassociation,
which implies a difficulty in treating the strong feedback mechanism with this model~\cite{Okada2}.
On the other hand, sparse coding for sequential associative memory
has not yet been studied in detail~\cite{Domany,Amari}.
In the present paper,
we study this point by applying the method of statistical neurodynamics
to a model for sequential associative memory.

Let us consider the situation in which a neural network
which consists of $N$ McCulloch-Pitts neurons
is designed to store sequential patterns rather than static ones.
Each neuron obeys discrete {\it synchronous} dynamics described by
\begin{eqnarray}
S_{i}(t+1)&=&F[h_{i}(t)] \label{eq:neuron}\\
h_{i}(t)&=&\sum_{j=1}^{N}J_{ij}S_{j}(t) -\theta \label{eq:localfield},
\end{eqnarray}
where $S_{i}(t)$ and $h_{i}(t)$ are the state and the internal potential of the $i$th neuron at time $t$, respectively.
Although we have written the transfer function in the general form $F(u)$,
we consider the case $F(u)=\Theta(u)$; i.e. $F(u)$ is a step function.
In this case, the state $S_{i}(t)$ takes only two values, 1 (firing state) and 0 (resting state).
The quantities $\theta$ and $J_{ij}$ represent the uniform threshold and the strength of the synaptic connection between the $i$th and $j$th neuron, respectively.

We assume that the stored patterns are generated with the probability $P(\xi_{i}^{\mu})=a\delta(\xi_{i}^{\mu}-1)+(1-a)\delta(\xi_{i}^{\mu})$.
Here $\xi_{i}^{\mu}$ is the state of the $i$th neuron in the $\mu$th pattern.
Then, the activity for this network, $\frac{1}{N}\sum_{i}\xi_{i}^{\mu}$,
assumes an average value of $a$.
In particular, the case $a \to 0$ is referred to as ``sparse coding''.
In order to make the network possess associative memory dealing with these patterns,
the $J_{ij}$s must be designed appropriately.
In the present paper,
to construct a network capable of recalling a sequence of $\alpha N$ patterns
defined by such as $\xi^{1} \to  \xi^{2} \to \cdots \to \xi^{\alpha N} \to \xi^{1} \to \cdots $,
we adopt covariance learning
\begin{equation}
J_{ij}=\frac{1}{a(1-a)N}\sum_{\mu}^{\alpha N}
(\xi_{i}^{\mu+1}-a)(\xi_{j}^{\mu}-a) \label{eq:synapse},
\end{equation}
which is usually adopted in the context of  learning the sparsely coded patterns.

For such a network, the macroscopic state is found
to be described by the following order parameters:
\begin{eqnarray}
m^{\mu}(t)&=& \frac{1}{a(1-a)N}\sum_{j}^{N}(\xi_{j}^{\mu}-a)S_{j}(t) \label{eq:defoverlap}\\
x(t)&=&\frac{1}{aN}\sum_{j}^{N}S_{j}(t) \label{eq:defactivity}.
\end{eqnarray}
Here, $m^{\mu}(t)$ is the overlap with the target pattern $\xi^{\mu}$.
As the configuration of the network becomes close to the target pattern,
this value approaches unity.
The function $x(t)$ represents the activity of the network.
On studying the retrieval processes,
we mainly discuss the time evolution of these parameters.

From this point, we consider the ``condensed'' situation
in which only one overlap is sizable: $m^{\rho}(t) \sim O(1)$, and $m^{\mu}(t) \sim O ( 1/\sqrt{N} )(\mu \ne \rho)$.
Here, $\xi^{\rho}$ is the pattern to be retrieved at time $t$.
Then, the internal potential $h_{i}(t)$ in (\ref{eq:localfield})
can be separated as
\begin{equation}
h_{i}(t)=\bar{\xi}_{i}^{\rho+1}m^{\rho}(t) -\theta 
+\frac{1}{a(1-a)N}\sum_{j \ne i}^{N}\sum_{\mu \ne \rho}^{\alpha N}
\bar{\xi}_{i}^{\mu+1}\bar{\xi}_{j}^{\mu}S_{j}(t),
\label{eq:sn}
\end{equation}
where we have written $\bar{\xi}_{i}^{\mu}$ as $\xi_{i}^{\mu}-a$.
In this process, the first and the second terms in (\ref{eq:sn}) are together
regarded as the signal to induce recollection of the target pattern $\xi_{i}^{\rho+1}$
at the subsequent time step, $t+1$, while the remaining term is regarded as noise.
For convenience, we define the noise term $z_{i}(t)$ as
\begin{equation}
z_{i}(t)=\frac{1}{a(1-a)N}\sum_{j \ne i}^{N}\sum_{\mu \ne \rho}^{\alpha N}
\bar{\xi}_{i}^{\mu+1}\bar{\xi}_{j}^{\mu}S_{j}(t)
\end{equation}
The quantity $z_{i}(t)$ is the crosstalk noise from the non-target patterns.
The essence of the theory is to treat the crosstalk noise $z_{i}(t)$
as Gaussian noise with mean $0$ and variance $\sigma(t)^{2}$~\cite{Amari}.
It has been confirmed numerically that this assumption is valid as long as the network succeeds in retrieval~\cite{Nishimori}.

Now we derive the dynamical equations for the overlap $m(t)$ and the activity $x(t)$.
The definition of the overlap leads to the equation
\begin{equation}
m^{\rho+1}(t+1)=\frac{1}{a(1-a)N}\sum_{i}^{N} \dlangle \bar{\xi}_{i}^{\rho+1}
F\left[ \bar{\xi}_{i}^{\rho+1}m^{\rho}(t)-\theta+z_{i}(t) \right] 
\drangle_{\xi},
\end{equation}
where $\dlangle \cdots \drangle_{\xi}$
denotes the average over the stored patterns.
In the same way, we can write the equation for the activity $x(t)$,
\begin{equation}
x(t+1)=\frac{1}{aN}\sum_{i}^{N} \dlangle
F\left[ \bar{\xi}_{i}^{\rho+1}m^{\rho}(t)-\theta+ z_{i}(t) \right]
\drangle_{\xi}.
\end{equation}

Next, we examine the time development of the variance $\sigma(t)^{2}$.
Expressing $z_{i}(t+1)$ as
\begin{equation}
z_{i}(t+1)=\frac{1}{a(1-a)N}\sum_{j \ne i}^{N}\sum_{\mu \ne \rho+1}^{\alpha N}
\bar{\xi}_{i}^{\mu+1}\bar{\xi}_{j}^{\mu}F[h_{j}(t)] \label{eq:noise},
\end{equation}
we must consider the dependence of $h_{j}(t)$ on $\xi_{j}^{\mu}$
when summing over $\mu$.
In the internal potential $h_{j}(t)$,
the term $\bar{\xi}_{j}^{\mu}m^{\mu-1}(t)$
is estimated to be $O(1/\sqrt{N})$.
Therefore, we expand the function $F[h_{j}(t)]$, obtaining
\begin{eqnarray}
z_{i}(t+1)&=&\frac{1}{a(1-a)N}\sum_{j \ne i}^{N}\sum_{\mu \ne \rho+1}^{\alpha N}
\bar{\xi}_{i}^{\mu+1}\bar{\xi}_{j}^{\mu}\hat{S}_{j}(t+1)+U(t)z_{i}(t), \label{eq:recursion} \\
U(t)&=& \frac{1}{a(1-a)N}\sum_{j}^{N} \dlangle
(\bar{\xi}_{j}^{\mu})^{2}F'[h_{j}(t)] \drangle_{\xi} \label{eq:def_U}.
\end{eqnarray}
We now assume that $\hat{S}_{j}(t+1)$ in (\ref{eq:recursion}) is independent of $\xi_{j}^{\mu}$.
Squaring (\ref{eq:recursion}), we obtain
\begin{eqnarray}
z_{i}(t+1)^{2}&=&\alpha a x(t+1) +U(t)^{2}z_{i}(t)^{2} \nonumber \\
&+&U(t) \frac{1}{a^{2}(1-a)^{2}N^{2}} \sum_{j,k}\sum_{\mu,\nu}
\bar{\xi}_{i}^{\mu+1} \bar{\xi}_{i}^{\nu+1}
\bar{\xi}_{j}^{\mu}\hat{S}_{j}(t+1)
\bar{\xi}_{k}^{\nu-1}S_{k}(t). \label{eq:square}
\end{eqnarray}
Here the first term and the second term in (\ref{eq:square})
come from the square of the first term and the second term in (\ref{eq:recursion}), respectively.
The last term in (\ref{eq:square}) arises from
the product of the first term and the second term in (\ref{eq:recursion}).
For the same reason, the term $S_{k}(t)$ in (\ref{eq:square}) must be expanded.
Following this procedure iteratively, we can take into account temporal correlations up to the initial time.
Averaging (\ref{eq:square}), the equation for $\sigma(t)$ takes
\begin{equation}
\sigma(t+1)^{2}=\alpha a x(t+1)+U(t)^{2}\sigma(t)^{2}
+\sum_{n=1}^{t+1}C(t+1,t+1-n)
\end{equation}
with
\begin{equation}
C(t+1,t+1-n)=\prod_{\tau=1}^{n}U(t+1-\tau)
\frac{1}{N}\sum_{j}\hat{S}_{j}(t+1)\hat{S}_{j}(t+1-n)
\left[ \frac{1}{a^{2}(1-a)^{2}N}\sum_{\mu}
(\bar{\xi}_{i}^{\mu+1})^{2}\bar{\xi}_{j}^{\mu}\bar{\xi}_{j}^{\mu-n} \right].
\end{equation}
Since $\xi^{\mu}$ and $\xi^{\mu-n}$ are independent of each other,
except when $n=\alpha N, 2\alpha N, 3\alpha N, \cdots$,
the last summation in the above equation vanishes.
Although the correlations for $n=\alpha N, 2\alpha N, 3\alpha N, \cdots$ remain,
their effect can be regarded as negligible in the limit $N \to \infty$.
Finally, we obtain
\begin{equation}
\sigma(t+1)^{2}=\alpha a x(t+1)+U(t)^{2}\sigma(t)^{2}.
\end{equation}
This derivation is essentially equivalent to that by Amari~\cite{Amari}.

Consequently, the behavior of the network is described by the equations
\begin{eqnarray}
m(t+1)&=&1-\frac{1}{2}\left[\erfc(\phi_{1})+\erfc(\phi_{0}) \right] \\
x(t+1)&=&1-\frac{1}{2}\left[\erfc(\phi_{1})-\frac{1-a}{a}\erfc(\phi_{0}) \right] \\
\sigma(t+1)^{2}&=&\alpha a x(t+1)+\frac{1}{2\pi}\left[ a\exp(-\phi_{1}^{2})+(1-a)\exp(-\phi_{0}^{2}) \right]^{2} \label{eq:sig_result},
\end{eqnarray}
with
\begin{eqnarray}
\phi_{1}&=&\frac{(1-a)m(t)-\theta}{\sqrt{2}\sigma(t)} \label{eq:phi1}\\
\phi_{0}&=&\frac{am(t)+\theta}{\sqrt{2}\sigma(t)} \label{eq:phi0},
\end{eqnarray}
where we have set $F(u)=\Theta(u)$ and 
replaced the site average $\frac{1}{N}\sum_{i}^{N} \cdots $ 
with the average over the Gaussian noise $\left< \cdots \right>_{z(t)}$
in the limit $N \to \infty$.
For initial values, we can set $\sigma(0)=\sqrt{\alpha a x(0)}$ and choose arbitrary values for $m(0)$ and $x(0)$.

In a sparsely coded network,
activity control is an important factor for good retrieval quality.
Introducing the global inhibitory interaction such as
\begin{equation}
J^{{\rm inh}}_{ij}=J_{ij}-\frac{g}{aN},
\end{equation}
the activity can be dynamically controlled~\cite{AGS,Buhmann}.
The second term contributes as a global inhibitory interaction,
and $g$ represents its strength.
If the activity level of the network at time $t$, $x(t)$, greatly increases,
each neuron receives a stronger inhibitory signal $-g x(t)$,
so that $x(t+1)$ decreases.
We can undertake a treatment of the retrieval process in this case
in a manner similar to that undertaken above.
We then find that equations (\ref{eq:phi1}) and (\ref{eq:phi0}) are modified as
\begin{eqnarray}
\phi_{1}&=&\frac{(1-a)m(t)-gx(t)-\theta}{\sqrt{2}\sigma(t)} \\
\phi_{0}&=&\frac{am(t)+gx(t)+\theta}{\sqrt{2}\sigma(t)}.
\end{eqnarray}

Another model possessing an activity control mechanism is
that with a time-dependent threshold
which is calculated at each time step
so that the activity of the network can be kept 
the same as that of the retrieved pattern~\cite{Okada2}.
Recently, as an improved model,
a ``self-control'' model has been proposed~\cite{Dominguez}.
In this model, the time-dependent threshold $\theta(t)$ adapts itself
according to the activity $a$ and the variance of crosstalk noise $\sigma(t)$.
If $a$ is sufficiently small, it takes the form $\theta(t)=\sigma(t)\sqrt{-2\ln a}$.
However, from the biological point of view, it is not plausible that
the network monitors the statistical quantity of the crosstalk noise.
Hence, in the present paper, in place of $\sigma(t)$,
we choose the leading term of $\sigma(t)$, $\sqrt{a \alpha x(t)}$.
Then, we simply use 
\begin{equation}
\theta(t)=\sqrt{-2 x(t) \alpha a \ln a}
\end{equation}
in place of the original expression.

We now compare our theoretical results with numerical simulations.
Figures \ref{fig:sim1}-\ref{fig:sim3} display the results
of the model using only a uniform threshold $\theta$,
a uniform threshold $\theta$ and the inhibitory interaction $g$,
and a self-control threshold $\theta(t)$, respectively.
In the first two cases,
$\theta$ and $g$ are optimized so as to maximize the storage capacity.
From the results, it is found that the theoretical curves provide
a good prediction of the retrieval properties in the network.
Although with respect to storage capacity, these three cases differ very little, the differences among the activity control methods are reflected in the shapes of basin of attraction.
While the basin becomes gradually narrow
as $\alpha$ increases in the first case,
the basin for $\alpha>0$ is wider than that for $\alpha=0$ in the second case.
Furthermore, in the last case,
the minimum initial overlap for which the network succeeds in retrieval becomes zero when $\alpha=0$.

Next, we investigate robustness against random synaptic dilution.
In this case, a randomly diluted synapse is 
represented by the random variable $c_{ij}$:
\begin{equation}
\tilde{J}_{ij}=\frac{c_{ij}}{c}J_{ij}
\end{equation}
The variable $c_{ij}$ takes the value 1 with probability $c$, and is 0 otherwise.
In other words, $c$ represents the ratio of connected synapses.
It is known that random synaptic dilution can be statistically regarded 
as static noise in a synapse~\cite{Sompolinsky},
and ultimately plays the role of a static noise, additional to the crosstalk noise in the retrieval dynamics~\cite{we}.
Therefore, the resultant equation for the noise is modified as
\begin{equation}
\tilde{\sigma}(t+1)^{2}=\sigma(t+1)^{2}+ a\alpha\frac{1-c}{c}. \label{eq:noise_w_dilution}
\end{equation}
The last term in (\ref{eq:noise_w_dilution}) is 
attributed to the  variance of synaptic noise caused by dilution.
In addition, $\phi_{1}$ and $\phi_{0}$ become
\begin{equation}
\phi_{1}=\frac{(1-a)m(t)-\theta}{\sqrt{2}\tilde{\sigma}(t)}
\end{equation}
\begin{equation}
\phi_{0}=\frac{am(t)+\theta}{\sqrt{2}\tilde{\sigma}(t)}.
\end{equation}

In order to examine the deterioration experienced
with the decrease in the ratio of connection $c$ at each activity $a$,
we define the normalized storage capacity
$\alpha_{c}^{\ast}(c)=\alpha_{c}(c)/\alpha_{c}(1)$,
where $\alpha_{c}(c)$ is the storage capacity when the ratio of connection is $c$.
(\ref{fig:dilution}) displays the normalized storage capacity $\alpha_{c}^{\ast}(c)$ as a function of $c$.
As indicated by these results, even if the activity level $a$ becomes small,
the shape of the curve representing the degree of deterioration does not change significantly.
However, as $a$ becomes small,
the storage capacity comes to decrease almost linearly
with the increase in the degree of dilution, $1-c$.
With regard to the basin of attraction, 
the model with the optimized uniform threshold $\theta$, 
which has the most narrow basin, is the most robust of the three.

Finally, we briefly mention the dependence of the storage capacity on the activity level $a$.
Also, in the present case, we have numerically confirmed that the storage capacity diverges as $-1/a \ln a$ in the limit $a \to 0$,
and it seems to approach such an asymptotic form quite slowly~\cite{HBFKS}.

We express our gratitude to M. Okada and T. Fukai for helpful comments
and Professor T. Munakata for valuable discussions.
This work is supported by the Japanese Grant-in-Aid for Science Research Fund
from the Ministry of Education, Science and Culture.

\newpage
\begin{figure}
\centering
\epsfile{file=figure1,scale=0.5}
\caption{
From the top, the equilibrium activity (dashed curve), equilibrium overlap (dotted curve), and basin of attraction (full curve)
for $a=0.1$ and $\theta=\theta_{opt}(=0.47)$.
The ordinate is the overlap $m$ or the activity $x$,
and the abscissa is the loading rate $\alpha$.
The data points indicate simulation results with $N=2000$ for 20 trials.
We take the initial activity as $x(0)=1.0$.
The inset shows the dependence of the storage capacity $\alpha_{c}$ on the uniform threshold $\theta$.
The value at the peak of the curve corresponds to $\theta_{opt}$.
}
\label{fig:sim1}
\end{figure}

\begin{figure}
\centering
\epsfile{file=figure2,scale=0.5}
\caption{
A plot similar to that in figure 1
for the case $\theta=0$ and $g=g_{opt}(=0.56)$.
The other parameters are the same as in figure 1.
The inset shows the dependence of the storage capacity $\alpha_{c}$
on the inhibitory interaction $g$ when $\theta=0$.
The value at the peak of the curve corresponds $g_{opt}$.
}
\label{fig:sim2}
\end{figure}

\begin{figure}
\centering
\epsfile{file=figure3,scale=0.5}
\caption{
A plot similar to that in figure 1 for the self-control model.
The other parameters are the same as in figure 1.
}
\label{fig:sim3}
\end{figure}

\begin{figure}
\centering
\epsfile{file=figure4,scale=0.5}
\caption{
Dependence of the normalized storage capacity $\alpha_{c}^{\ast}(c)$
on the ratio of connected synapses $c$ 
for $a=0.5, 0.1, 0.001$.
}
\label{fig:dilution}
\end{figure}

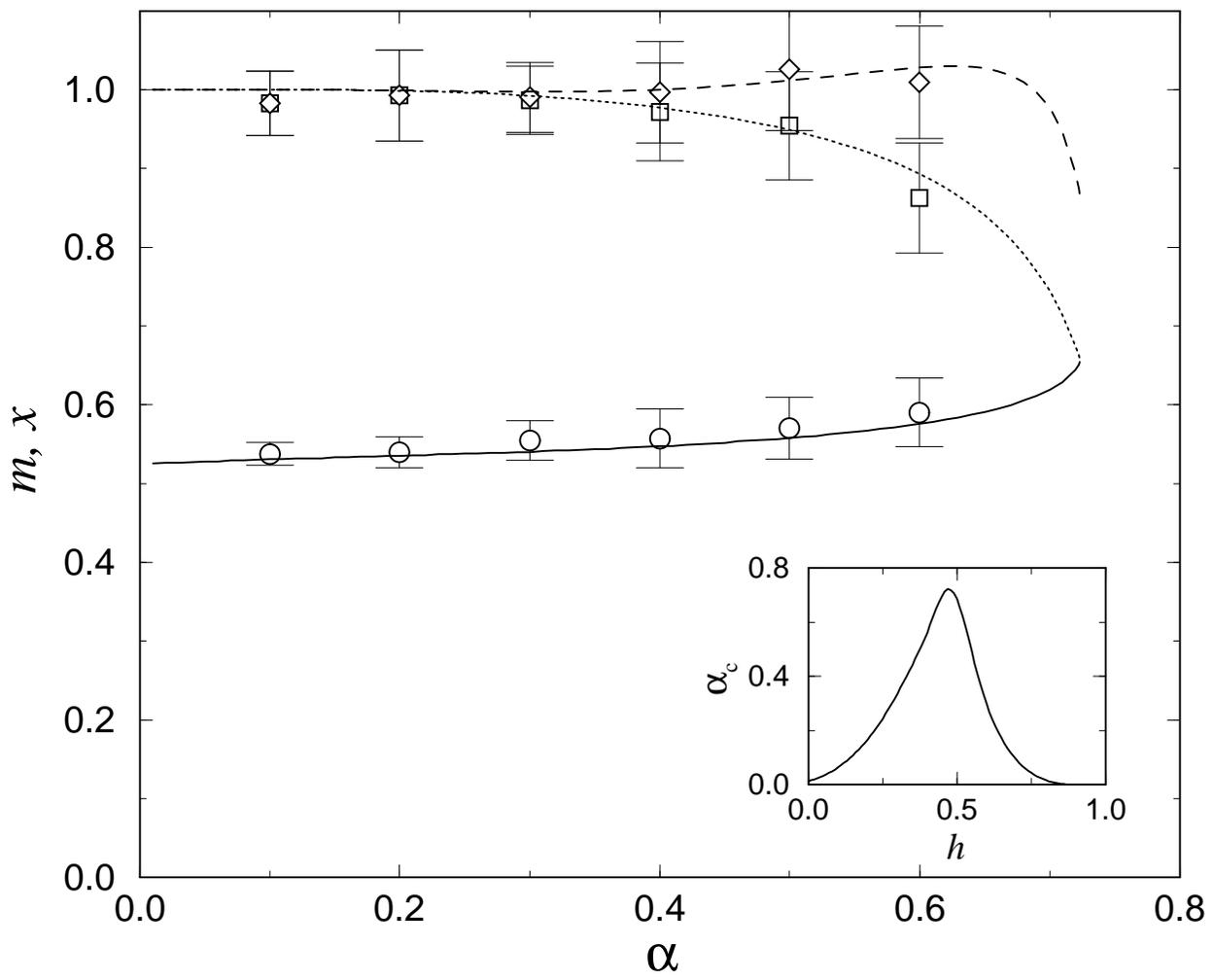
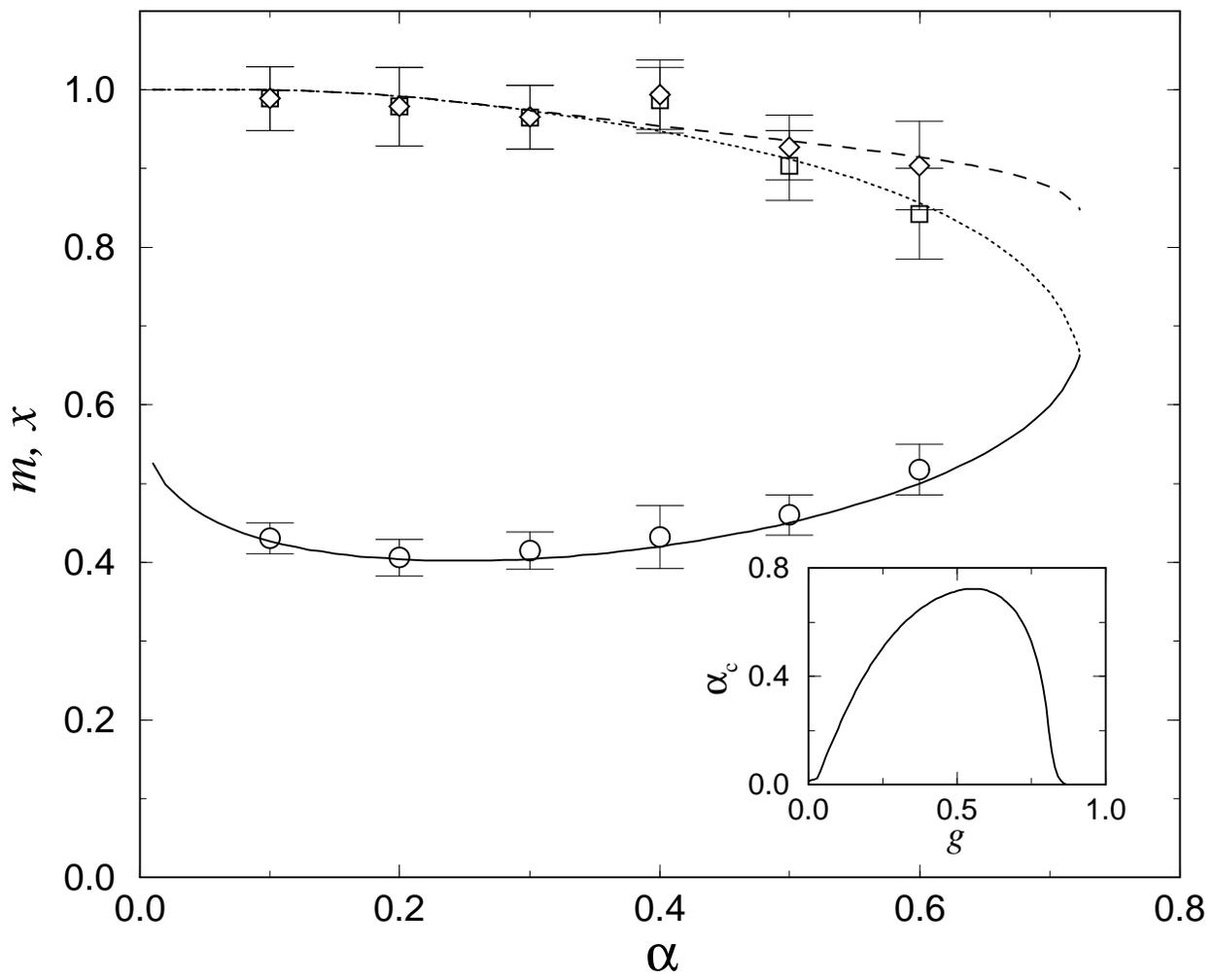
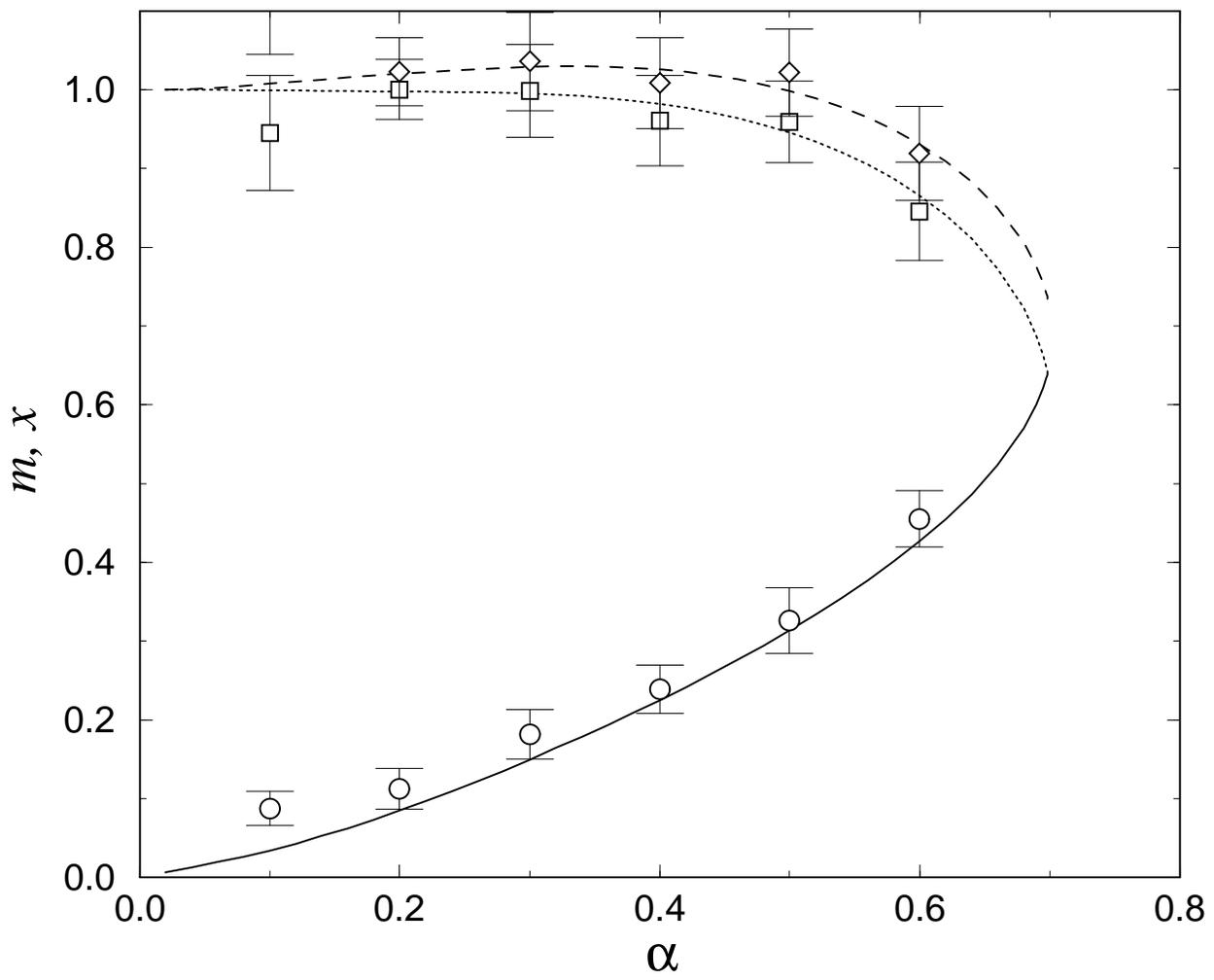
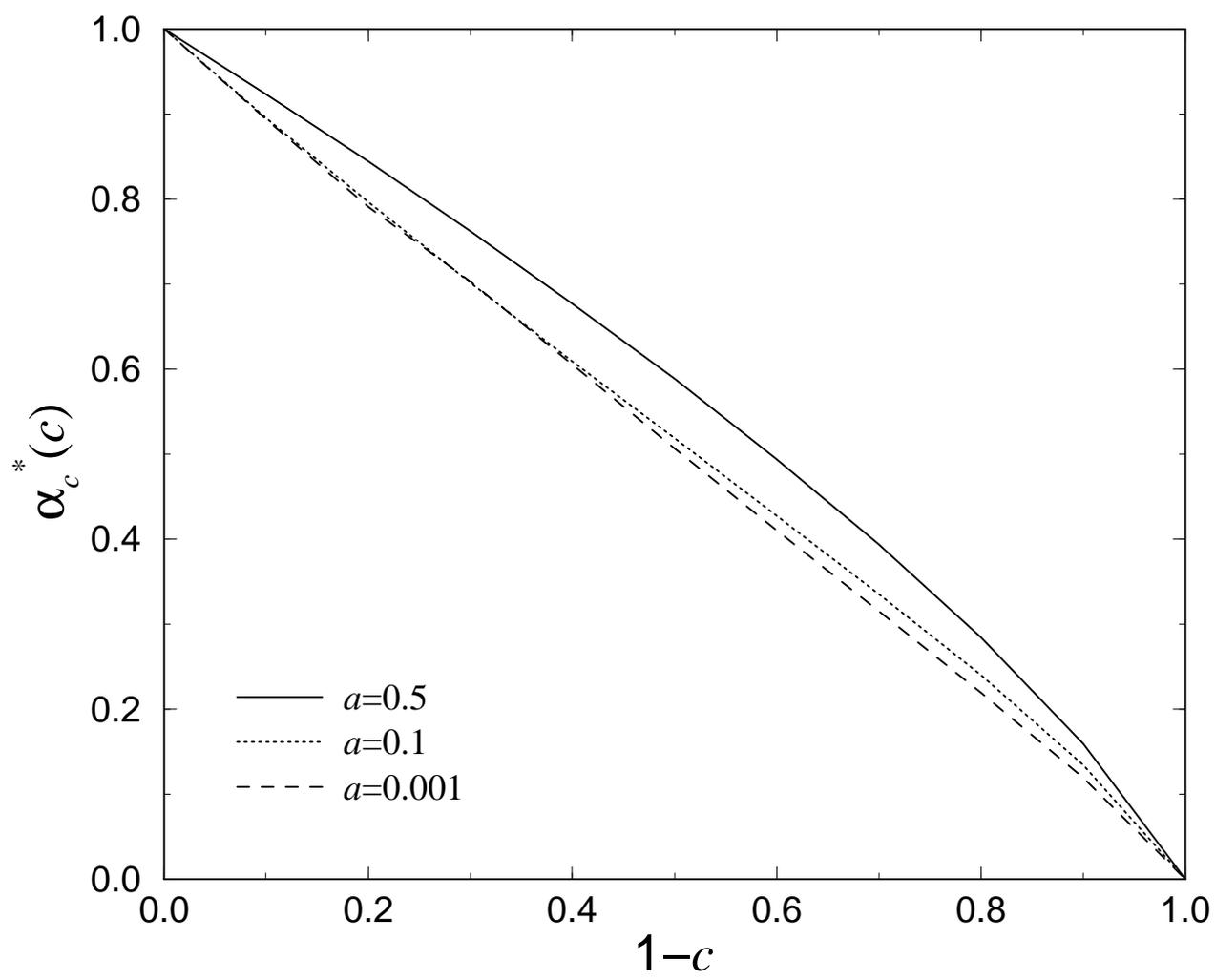
\end{document}